# Single-cycle THz pulses with amplitudes exceeding 1 MV/cm generated by optical rectification in LiNbO$_3$

H. Hirori[1,2], A. Doi[2,3], F. Blanchard[1,2], and K. Tanaka[1,2,4]

[1] *Institute for Integrated Cell-Material Sciences, Kyoto University, Sakyo-ku, Kyoto 606-8501, Japan*
[2] *CREST, Japan Science and Technology Agency, Kawaguchi, Saitama 332-0012, Japan*
[3] *Olympus Corporation, Hachioji-shi, Tokyo 192-8512, Japan*
[4] *Department of Physics, Graduate School of Science, Kyoto University, Sakyo-ku, Kyoto 606-8502, Japan*

Using the tilted-pulse-intensity-front scheme, we generate single-cycle terahertz (THz) pulses by optical rectification of femtosecond laser pulses in LiNbO$_3$. In the THz generation setup, the condition that the image of the grating coincides with the tilted-optical-pulse front is fulfilled to obtain optimal THz beam characteristics and pump-to-THz conversion efficiency. The designed focusing geometry enables tight focus of the collimated THz beam with a spot size close to the diffraction limit, and the maximum THz electric field of 1.2 MV/cm is obtained.

Recent successful developments in efficient high-power terahertz (THz) pulse generation has created many promising applications such as in large-scale object imaging, medical diagnosis and treatment, and remote sensing techniques for security issues.[1,2] In addition, intense THz pulses allow for study of unexplored nonlinear phenomena such as coherent THz manipulation of electric and magnetic quantum states,[3,4,hub] high harmonics generation,[5] nonlinear optical processes, [6-11,Z-scan] and nonlinear transport phenomena in solids. [12]

Since Hebling *et al.* (2002) proposed a tilted-pump-pulse-front scheme for efficient phase-matched THz pulse generation using $LiNbO_3$ crystals,[13] the technique has been rapidly developing. The technique has demonstrated the possibility of THz pulse generation with energies on the scale of 10 μJ.[14,15] To make the technique versatile for applications and useful for study of unexplored nonlinear phenomena, a generation setup to obtain optimal THz beam characteristics and a maximized THz peak field is required. A recent detailed analysis of the scheme predicted that the imaging errors in the setup consisting of a grating and lenses can lead to distortion in THz intensity profile after the $LiNbO_3$ output surface.[16] This distortion could create strong and ambiguous divergence in the THz beam, causing inaccuracy in optimal THz measurement geometry design, thereby limiting its applications.

In this paper, we report the generation of single-cycle THz pulses using the tilted-pump-pulse-front scheme with a 1.3 mol% MgO-doped stoichiometric $LiNbO_3$ (LN) crystal. In the THz generation setup, the condition that the image of the grating coincides with the tilted-optical-pulse front is fulfilled to obtain optimal THz beam characteristics and pump-to-THz conversion efficiency. The propagation characteristics of the THz beam generated from the LN crystal were measured by an uncooled microbolometer-array THz camera. The results show that the THz beam had divergence of 52 ± 5 mrad in the horizontal direction for 1 THz. The designed focusing geometry for the collimated THz beam enables tight focus onto the electro–optic (EO) crystal with a spot size of around 300 μm for 1 THz, and a peak THz electric field of 1.2 MV/cm and energy conversion efficiency of ~1 × 10$^{-3}$ are obtained.

Figures 1(a) and 1(b) show schematics of the THz pulse generation setup using the tilted-pump-pulse-front scheme and the EO sampling setup, respectively. For a pump source, we used an amplified Ti:sapphire laser that provided an average power of 4 mJ, a full width at half-maximum intensity (FWHM) of 85 fs, a central wavelength of 780 nm, and a repetition rate of 1 kHz. The optical pulses from the oscillator with a repetition rate of 80 MHz synchronized with the amplified pulses were used for the EO sampling.

To match the noncollinear velocity of the pump $v_p^{gr}$ and THz pulses $v_{THz}$ in the LN crystal for efficient THz generation, i.e., $v_p^{gr} \cos\gamma = v_{THz}$, we tilted the pump-pulse front angle $\gamma$ by using the grating and two cylindrical lenses, as shown in Fig. 1(a). The angle $\gamma$ is described as below:[16]

$$\tan\gamma = \frac{m\lambda_0 p}{n_p^{gr} \beta_1 \cos\theta_d}, \qquad (1)$$

where $m$, $p$, and $\theta_d$ are the diffraction order, diffraction angle, and groove density of a grating, respectively. $\lambda_0$ and $n_p^{gr}$ are the central wavelength and group refractive index of the LN crystal for the pump pulse, respectively. $\beta_1$ is the horizontal magnification factor of the lenses for the pump-pulse front, which is given by a ratio of $f_2/f_1$. The LN crystal angle $\theta_{LN}$ in Fig. 1 (a) should be the same as the $\gamma$ for the THz pulse to propagate in the normal direction to the LN output surface.

To obtain optimal THz beam characteristics and pump-to-THz conversion efficiency, the tilt angle inside the LN of the grating image $\theta$ should coincide with that of the pump-pulse front $\gamma$ because the temporal pump-pulse duration across the image of the grating is minimal. The tilt angle $\theta$ is described as follows:[16]

$$\tan \theta = n \beta_2 \tan \theta_d \qquad , \qquad (2)$$

where $n$ is the refractive index of the LN crystal for the pump pulse. $\beta_2$ is the horizontal magnification factor of the lenses for the grating image, which is also given by a ratio of $f_2/f_1$.

Figure 2 shows the calculated curves of the horizontal magnification factors $\beta_1$ for the different grating grooves and $\beta_2$ as functions of $\theta_d$.[17] Here, $\theta$, $\gamma = 62°$ in Eqs. (1) and (2) are set for efficient generation of around 1 THz.[18] When the magnification factors have the same value, the condition of matching the angles, i.e., $\theta = \gamma = 62°$, is fulfilled. As shown in Fig. 2, the curve $\beta_1$ for the grating groove of 1800 cm$^{-1}$ (solid line) and $\beta_2$ (dashed line) have the same value of 0.59 when $\theta_d = 55.7°$ as indicated by an arrow. In the actual setup, the horizontal magnification factor of $f_2/f_1 = \beta_1 = \beta_2 = 0.6$, which is close to the calculated ideal value of 0.59, is realized by using 4$f$-lens geometry as shown in Fig. 1(a). The 4$f$-lens geometry may be suitable for the generation of the collimated THz beam.

Figures 3(a) and 3(b) show the THz intensity image and cross-section profiles at 45 mm from the LN output surface, respectively. The THz intensity image passing through a 300 GHz width band-pass filter for 1 THz (Murata Manufacturing Co., Ltd., MMBPF40-1000), can be measured by a 320 × 240 pixel uncooled microbolometer THz camera with 23.5 μm pixel pitch from NEC Corporation (model IRV-T0830). In Fig. 3(b), the solid lines show the corresponding Gaussian fits, and diameters at FWHM in vertical and horizontal directions are 1.9 mm and 1.3 mm, respectively. The image is vertically elongated because the incident pump pulse with a spot size of 5.2 mm at FWHM was horizontally magnified with a factor of 0.6.

Figure 3(c) shows the spot size as a function of propagation distance. The THz beam has a divergence of 54 ± 5 and −5 ± 5 mrad in horizontal and vertical directions, respectively. The fairly collimated beam with radially symmetric Gaussian beams shown in Figs. 3(c) and 3(b) indicates that the THz generation optics with 4$f$-lens geometry can make the angle of the grating image coincide with that of the pump-pulse front. Otherwise, the spatially asymmetric intensity profile and strong divergence of THz beam could be induced.[16]

To obtain the maximized peak THz field, the generated collimated THz beam should be expanded, collimated, and focused tightly. As shown in Fig. 1(b), we developed the focusing geometry onto the EO crystal by three off-axis parabolic mirrors PM1, PM2, and PM3, assuming the collimated THz beam source rather than the point source.

Figures 4(a) and 4(b) show the THz temporal profile, measured by THz EO sampling with a 300 μm thick GaP detection crystal, and its Fourier components, respectively. In the EO detection, six high-resistivity Si attenuaters were inserted to reduce the field amplitude before the GaP detection crystal.[19] The maximum modulation of the balanced photodetector signals $I_A$ and $I_B$ measured at the peak THz field, i.e., $(I_A - I_B)/(I_A + I_B)$, is 0.44 with the Si attenuators; this value corresponds to the electric field of 1.2 MV/cm without the Si attenuators.[20] The spectrum in Fig. 1(b) has a maximum intensity around 1 THz and absorption lines caused by water vapor because dry air purging was not performed.

Figures 4(c) and 4(d) show the THz image and intensity profiles, respectively, at the focused point after the last off-axis parabolic mirror (PM3 in Fig. 1(b)). The intensity profiles in Fig. 4(d) show that the both THz spot diameters at FWHM in vertical and horizontal directions were around 300 μm. The small spot size close to the diffraction limit implies that our designed focusing optics shown in Fig. 1(b) directs a sufficiently tightly focused THz beam. The total pulse energy was estimated to ~2 μJ by integrating the THz intensity both temporally and spatially,[21] and the energy conversion efficiency was ~$1 \times 10^{-3}$. THz pulse energy obtained by a pyroelectric detector (MicroTech Instruments) indicates a larger value of ~3 μJ.

In conclusion, we generated single-cycle THz pulses using the tilted-pump-pulse-front scheme with LN crystals. In the THz generation setup, the condition that the image of the grating coincides with the tilted-optical-pulse front is fulfilled for the generation of collimated THz beam with an optimal pump-to-THz conversion efficiency. The designed focusing geometry, assuming inclusion of the collimated THz beam source rather than the point source, enabled us to tightly direct the THz beam with a spot size of ~300 μm for 1 THz, and the peak THz electric field of 1.2 MV/cm and energy conversion efficiency of ~$1 \times 10^{-3}$ were obtained.

We are grateful to Mitsuru Namiki for valuable discussions and also to Iwao Hosako and Kazuhiko Oda for letting us use the THz camera. This study was supported by a Grant-in-Aid for Scientific Research from JSPS (Grant No. 21760038) and from MEXT of Japan (Grant Nos. 18GS0208 and 20104007).

**Figure Captions**

**Fig. 1 (Color online).** (a) Schematic of the THz pulse generation setup for the tilted-pump-pulse-front scheme. The 4*f*-lens configuration consists of two cylindrical lenses L1 and L2 with respective focal lengths of $f_1$=250 mm and $f_2$=150 mm in horizontal direction. The grating we used has a groove density of 1800 cm$^{-1}$. The incident angle $\theta_i$ and diffracted angle $\theta_d$ of the grating are set at 35.3° and 55.7°, respectively. The LN prism angle $\theta_{LN}$ is 62°. (b) Schematic of the electro–optic sampling setup. The off-axis parabolic mirrors PM1, PM2, and PM3 have effective focal lengths of 10, 100, and 50 mm, respectively. Their respective diameters are 10, 50, and 50 mm. M: mirror, WP: Wollaston prism, PD: photo detector.

**Fig. 2 (Color online).** Calculated magnifications of the pump-pulse front $\beta_1$ for the different grating grooves and the grating image $\beta_2$ as a function of $\theta_d$ in Eqs. (1) and (2), respectively.

**Fig. 3 (Color online).** (a) THz intensity image and (b) its cross-section profiles with Gaussian fits measured at 45 mm from the LN output surface. (c) Measured spot size is a function of propagation distance. Solid lines show linear fits.

**Fig. 4 (Color online).** (a) Measured THz temporal profile and (b) its Fourier components. (c) The THz image measured at the focused point after the last off-axis parabolic mirror (PM3 in Fig. 1(b)). (d) Cross-section intensity profiles of the THz image in (c) with Gaussian fits.

**Fig. 1**

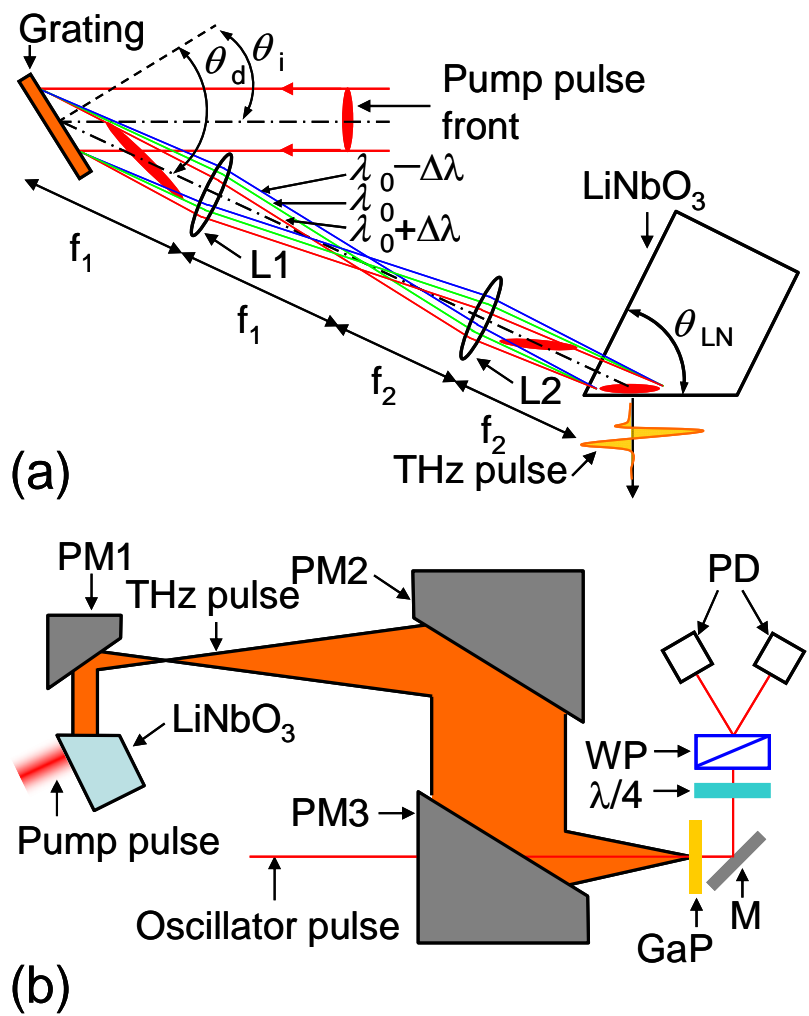

(a)

(b)

**Fig. 2**

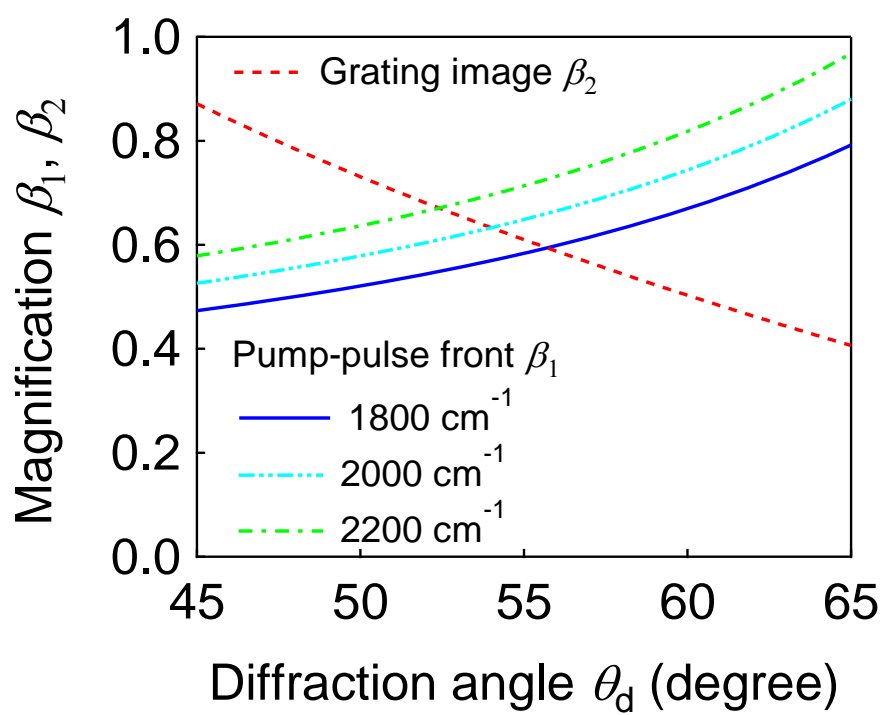

H. Hirori

**Fig. 3**

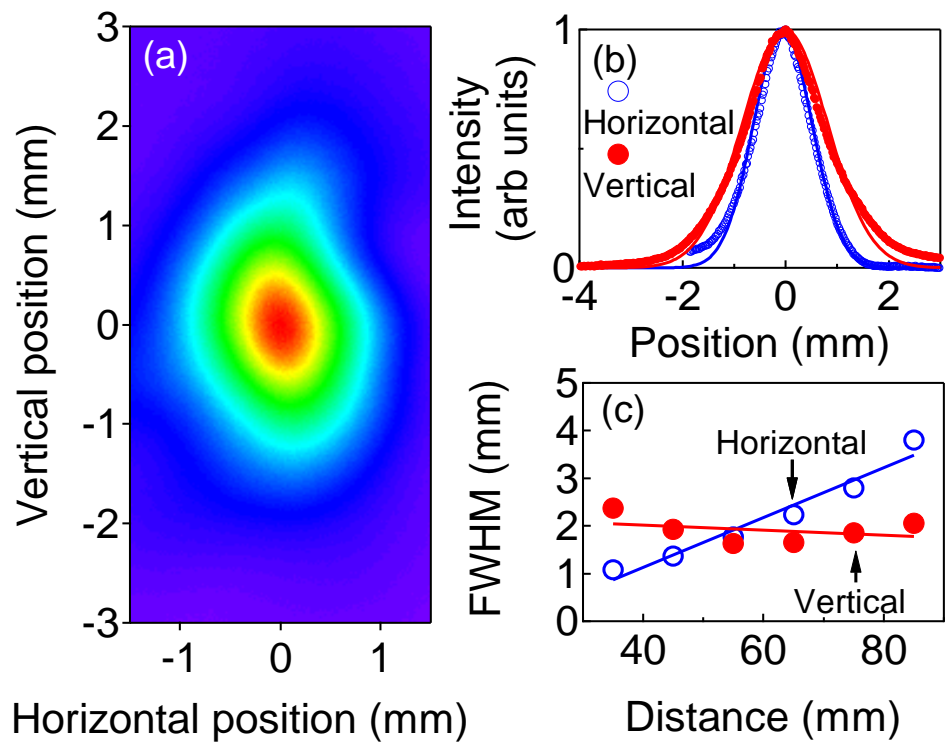

**H. Hirori**

Fig. 4

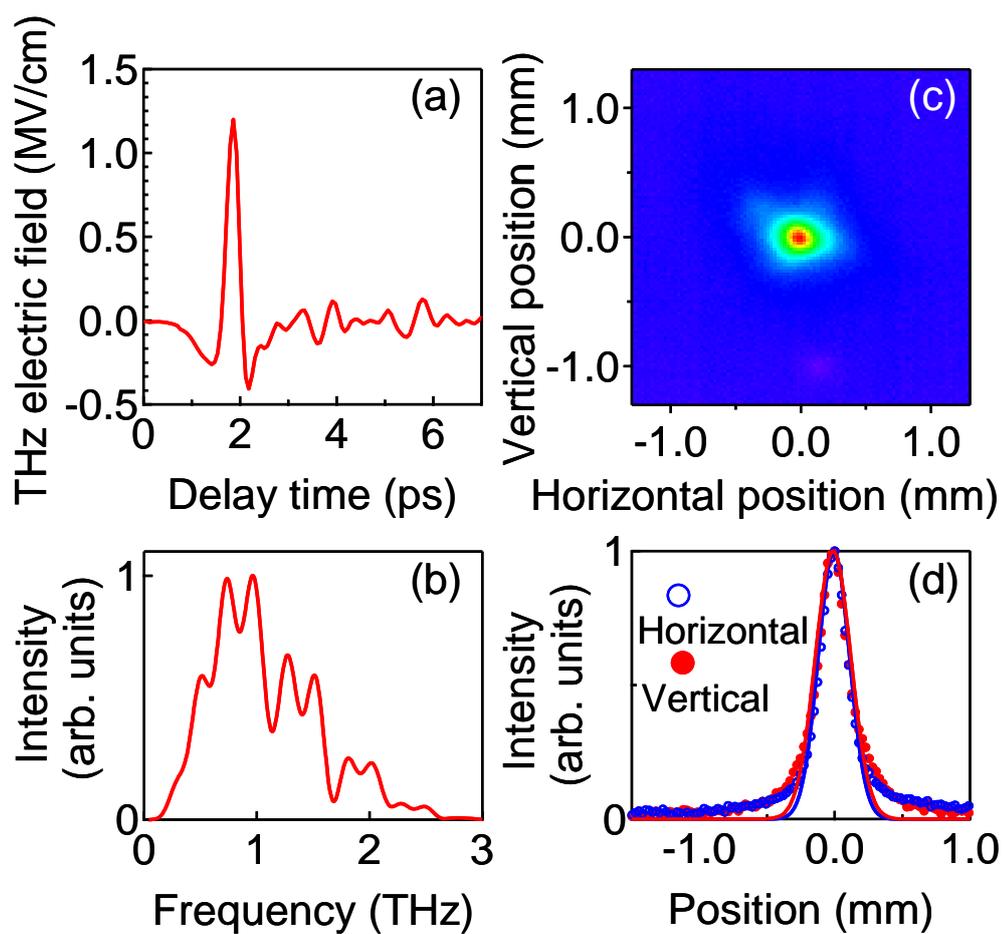

H. Hirori